\documentclass{elsart}

\input{psfig}
\usepackage{amssymb}

\begin{document}

\begin{frontmatter}



\title{Mass Number Dependence of Nuclear Pairing.}


\author{
S. Hilaire$^{1}$, J.-F. Berger$^{1}$, M. Girod$^{1}$, W. Satu{\l}a$^{2,3}$, 
and P. Schuck$^{4}$}

\address{$^1$D\'epartement de Physique Th\'eorique et Appliqu\'ee, CEA/DAM Ile-de-France, B.P. 12 - F91680 Bruy\`eres-le-Ch\^{a}tel, France}
\address{$^2$Institute of Theoretical Physics, University of Warsaw,
             ul. Ho\.za 69, PL-00-681 Warsaw, Poland}
\address{$^3$Royal Institute of Technology, Physics Department Frescati, 
             Frescativ\"agen 24, S-104 05 Stockholm, Sweden}
\address{$^4$Institut de Physique Nucl\'eaire, Universit\'e Paris--Sud,
             F-91406 Orsay Cedex, France}

\date{\today}

\maketitle

\begin{abstract}
 Large scale Hartree-Fock-Bogoliubov (HFB) calculations with the finite-range
 Gogny force D1S have been performed in order to extract the corresponding
 theoretical average mass dependence of the nuclear gap values. Good
 agreement with experimental data from the three-point filter
 $\Delta^{(3)}(N)$ with $N$ odd has been found for both the neutron
 and proton gaps. The results of our study support  
 earlier findings [W. Satu{\l}a, J. Dobaczewski, and W. Nazarewicz, 
 Phys. Rev. Lett. {\bf 81} 3599 (1998)]
 that the mass dependence of the gap is much weaker than the so far
 accepted $12 \, A^{-1/2}\,$ MeV law.
\end{abstract}

\end{frontmatter}


 In recent years, the study of pairing properties in systems of condensed
 matter so small that the coherence length of the Cooper pairs becomes 
 comparable with the size of the system has increased considerably. This is the 
 case for ultra small superconducting metallic grains \cite{braun} but one also 
 thinks that magnetically trapped fermionic atoms like $^{6}$Li can become
 superfluid at temperatures which may be reached experimentally in the near
 future \cite{marco}. A fermionic system of finite size where the superfluid 
 properties have been studied experimentally and theoretically since decades is
 the atomic nucleus \cite{bohr}. Very efficient mean field approaches have
 been developped in the past to account quantitatively for a great amount of
 experimental data. One of the most successfull models in this context is that
 developped by Gogny and collaborators with the use of a finite range effective
 interaction D1S \cite{Dec80,cpc91}. However, the global mass number (A)
 dependence of the gap has never been investigated in a systematic way using
 this force. Such a study has now become particularly timely because it has
 been observed \cite{Sat98} that the commonly accepted law for the average gap
 parameter ($\Delta=12 \, A^{-1/2}\,$ MeV) strongly overestimates the gap
 values in light nuclei. Empirical information concerning gap parameters can
 be derived in principle from large-scale analysis of odd-even staggering
 (OES) of nuclear binding energies. One should bear in mind, however, that
 there are two basic physical mechanisms behind OES, namely : (i) an effect of
 spontaneous breaking of spherical symmetry (Jahn-Teller mechanism
 \cite{Jah37} or shape effect) and, (ii) the blocking of pair correlation
 by an unpaired fermion. Determination of the pairing component of OES
 therefore requires a careful deconvolution, at least to the extent
 possible, of both effects. Thus the aim of this work is to demonstrate
 that the  average gap parameters at the Fermi surface deduced from large-scale
 unconstrained Gogny-HFB calculations are consistent with 
 component of OES deduced from empirical data according to the
 method proposed in \cite{Sat98}. In particular, it will be shown that
 the theoretical A dependence of the average gap is 
 much weaker than $12/\sqrt{A}$ dependence what is in 
 coincidence with the experimental data analysis performed in \cite{Sat98}.

 The simplest way to quantify the OES of binding energies is to use the
 three-point filter
 \begin{equation} \label{d3}
 \Delta^{(3)}(N) = 
 {\pi_N \over 2}  [B(N-1) + B(N+1) -2B(N)]
 \end{equation}
 where $\pi_N = (-1)^N$ is the number parity and $B(N)$ the (negative)
 binding energy of the system of particle number $N$. Eq.~(\ref{d3}) assumes
 proton number $Z$ to be fixed and thus provides neutron OES. An expression
 appropriate for proton OES can be obtained by replacing $N$ with $Z$ in
 Eq.~(\ref{d3}) and by fixing the neutron number $N$.

 In nuclear structure studies, filter (\ref{d3}) is not considered as an
 appropriate measure of the neutron or proton pairing gaps. This is mainly
 due to strong symmetry energy [$B_{sym}\propto (N-Z)^2$] contributions.
 However, because symmetry energy is number-parity independent and rather
 weakly depends on shell effects, its influence can be removed by using
 higher order filters like the four-point formula:
 \begin{equation}
 \Delta^{(4)}(N)=\frac{1}{2} [\Delta^{(3)}(N) + \Delta^{(3)}(N-1)]  \label{d4}.
 \end{equation}
 Global analysis of empirical data using filter (\ref{d4}) leads to the
 commonly used estimate $\Delta = 12 \, A^{-1/2}\,$ MeV for the pairing
 gap \cite{Zeldes}. This classical way of reasoning leading from formula
 (\ref{d3}) to (\ref{d4}) has its roots in the macroscopic-microscopic model.
 It assumes~\cite{Jen84} that the major contribution [apart from pairing]
 to (\ref{d3}) comes from the smooth liquid-drop component of the total
 energy (or more precisely from the symmetry energy term as mentioned above)
 while the shell-correction energy $\delta E_{shell}$ varies slowly enough
 with $N$ and $Z$ to neglect its contribution to (\ref{d3}) or (\ref{d4}).
 This assumption is, however, hardly acceptable because $\delta E_{shell}$
 is by definition the difference between the strongly oscillating
 shell-energy, $E_{sp}=\sum_{occup} e_i$, and the smooth Strutinsky-smeared
 energy, $\tilde{E}_{sp}$. The single-particle (sp) shell-energy term, $E_{sp}$,
 gives rise to OES which is well recognized in metallic clusters~\cite{Man94}.
 In the extreme case of independent particles (fermions) filling two-fold
 Kramers-degenerated levels of a fixed, deformed potential well, the sp OES
 is $\Delta^{(3)}_{sp}(2n+1) \approx 0$ and 
 $\Delta^{(3)}_{sp}(2n) \approx (e_{n+1}-e_n)/2$, where $e_n$ and $e_{n+1}$
 stand for {\it effective\/} Nilsson levels at the Fermi energy~\cite{Sat98}.

 In a previous study~\cite{Sat98}, it has been demonstrated, using
 self-consistent Skyrme-Hartree-Fock calculations, that the contribution
 to (\ref{d3}) due to the smooth Strutinsky energy, $\tilde{E}_{sp}$, nearly
 cancels out the contribution coming from the liquid-drop symmetry energy.
 Consequently, only $\Delta (N)\equiv\Delta^{(3)}(N=2n+1)$ can be considered
 as a probe of the {\it pairing\/} component of OES, while
 $\Delta^{(3)}(N=2n)$ mixes both mean-field and pairing effects. Note,
 that filter (\ref{d4}) always mixes pairing and sp components whatever the
 number-parity is.

 These ideas have recently been tested within a wide class of exactly
 solvable models invoking monopole pairing Hamiltonians ~\cite{Dob,gatl}.
 Although these models do always oversimplify various properties of complex
 nuclei, these studies clearly indicate the correctness of the proposed
 method, particularly for weak and intermediate pairing correlations, which
 is by far the most commonly encountered situation in finite nuclei.
 In this case a consistency between the BCS (or HFB) pairing gap and the
 $\Delta^{(3)}(2n+1)$ filter has been found as well.

 We therefore think that $\Delta^{(3)}(N=2n+1)$ is the best suited filter
 for the extraction of gap values from experimental data. 
 One should be aware, however, that there is ongoing debate 
 concerning detailed interpretation of $\Delta^{(3)}(N=2n+1)$
 as well as higher order filters [$\Delta^{(5)}$]~\cite{Bender,Dug1,Dug2}. 
 In particular, an effect of {\it time-odd\/} mean-field was intensively
 studied in Ref.~\cite{Dug2}. This mean-field effect 
 indeed enters directly empirical $\Delta^{(3)}(N=2n+1)$ 
 [as well as $\Delta^{(4,5)}$] through odd-A nuclei and should, 
 in principle, be removed explicitely. However, our knowledge concerning its
 magnitude is highly uncertain. For example, systematic 
 Skyrme-Hartree-Fock calculations of Ref.~\cite{gatl} indicate 
 attractiveness (repulsiveness) 
 of this effect for SLy4 
 (SIII, SkM$^*$) respectively with average absolute magnitude
 of the order of 100keV in odd-A light nuclei.


 Extensive HFB calculations have been performed in order to determine the
 ground state structure of nearly 400 even-even nuclei located in the
 neighborhood of those for which experimental pairing gaps have been
 extracted~\cite{Sat98}. The D1S parameterization of the Gogny
 Force~\cite{Dec80,cpc91} has been employed throughout this work. 
 Theoretical pairing gaps are then deduced from the pairing field obtained
 with this force in these nuclei, with the purpose of making comparisons
 with experimental gaps. It is of importance to point out that the calibration
 of the matrix elements of the Gogny force in the pairing channel has been
 based on OES in tin isotopes~\cite{Dec80} and that the A dependence of the
 calculated pairing gaps is ultimately governed by self-consistency
 requirements of the HFB solutions.

 According to the Bogoliubov theory~\cite{bogo}, the quasiparticle states
 can be obtained from the iterative diagonalization of the HFB Hamiltonian
 \begin{equation}\label{bog}
  H = \left( \begin{array}{cc} h -\mu I & -\Delta\\ -\Delta &  -h +\mu I
 \end{array}  \right) ,
 \end{equation}
 where $h$ and $\Delta$ are the matrices of the Hartree-Fock hamiltonian -- the
 sum of the nucleon kinetic energy and average field -- and of the pairing
 field in the HO basis, respectively, $\mu$ represents the chemical potential
 ensuring conservation of nucleon numbers, and $I$ is the unity matrix. Using
 time-reversal symmetry and appropriate phase conventions, $h$ and $\Delta$
 can be taken as real symmetric matrices.

 In order to derive from the HFB method theoretical quantities corresponding
 to empirical proton and neutron pairing gaps, the following technique is used.
 Single-particle energies  $\varepsilon_{i}$, pairing gaps $\Delta_{i}$ and
 occupation probabilities $v_{i}^2$ analogous to those defined in BCS theory
 are first calculated. They can be derived by either diagonalizing the
 Hartree-Fock Hamiltonian $h$, or expressing all relevant quantities in the
 canonical basis~\cite{BloMes}. In the first case, the $\varepsilon_{i}$ are
 taken as the eigenvalues of $h$, and the $\Delta_{i}$ and $v_{i}^2$ as the
 diagonal components of $\Delta$ and of the one-body density matrix $\rho$
 once they are expressed in the Hartree-Fock representation. In the second
 case, the  $v_{i}^2$ are the eigenvalues of $\rho$, while the $\Delta_{i}$
 and $\varepsilon_{i}$ are taken as the diagonal components of $\Delta$ and
 $h$ in the canonical basis. The two methods have been checked to yield very
 close single particle energies and practically identical values of the
 $v_{i}^2$ and $\Delta_{i}$~\cite{girod}. In the present work, the first
 method has been employed, and we will assume that the above quantities have
 their usual physical meaning. When applied separately to each kind of
 nucleons, single particle quantities denoted
 $\varepsilon_{i}^{\pi}$, $\Delta_{i}^{\pi}$, ${v_{i}^{\pi}}^2$ for protons,
 and $\varepsilon_{i}^{\nu}$, $\Delta_{i}^{\nu}$, ${v_{i}^{\nu}}^2$ for
 neutrons can thus be derived for all nuclei under consideration.

 Numbers representing the proton and neutron pairing gaps $\Delta^{\pi}$ and
 $\Delta^{\nu}$ in each nucleus have then been defined in two different ways.
 On the one hand, we define $\Delta_{last}^{\pi} = \Delta_{i=Z}^{\pi}$ and
 $\Delta_{last}^{\nu} = \Delta_{i=N}^{\nu}$, where  $i=Z$ (resp. $i=N$) is the
 $Z$th (resp. $N$th) proton (resp. neutron) state counted from the deepest one.
 On the other hand, we define 
 \begin{equation}
 \Delta_{aver}^{\pi}=
 \displaystyle{\sum_{i}} {u_{i}^{\pi}}^2 {v_{i}^{\pi}}^2 \Delta_{i}^{\pi} /
 \displaystyle{\sum_{i}} {u_{i}^{\pi}}^2 {v_{i}^{\pi}}^2
 \end{equation}
 and similarly for neutrons. In the second definition, the individual level
 gaps $\Delta_{i}$ are averaged out around the Fermi surface, with weights
 equal to the pair correlation probability of the two nucleons on each level.
 The purpose of this averaging is to smear out the sometimes large
 fluctuations ($\approx$ 100  keV) obtained for the individual $\Delta_{i}$'s
 in the vicinity of the Fermi surface. The gaps derived from the two methods
 are both functions of the nucleus proton and neutron numbers $Z$ and $N$.

 Finally, as in Ref.~\cite{Sat98}, proton (resp. neutron) pairing gaps averaged
 over $N$ (resp. $Z$) are defined as
 \begin{equation}\label{avera}
 \overline{\Delta_{type}^{\pi}} (Z) =
  { 1 \over M}\sum_{N=N_{1},N_{2},...,N_{M}} \Delta_{type}^{\pi} (N,Z).
 \label{defpmic}
 \end{equation}

 \noindent where $type$ is either $last$ or $aver$. We have decided to
 compare experimental $\Delta^{(3)}$'s with the pairing gaps of
 Eq.$(\ref{defpmic})$ instead of theoretical $\Delta^{(3)}$'s because the
 D1S Gogny force has not been designed to reproduce masses of odd-even or
 odd-odd nuclei. Indeed, HFB calculations do not account for particle-vibration
 coupling (which is known to be responsible for a decrease of a few hundreds of
 keV of the odd-even or odd-odd nuclei masses) but correctly describe even-even
 nuclei pairing properties.

\begin{figure}
\centerline{\hbox{\psfig{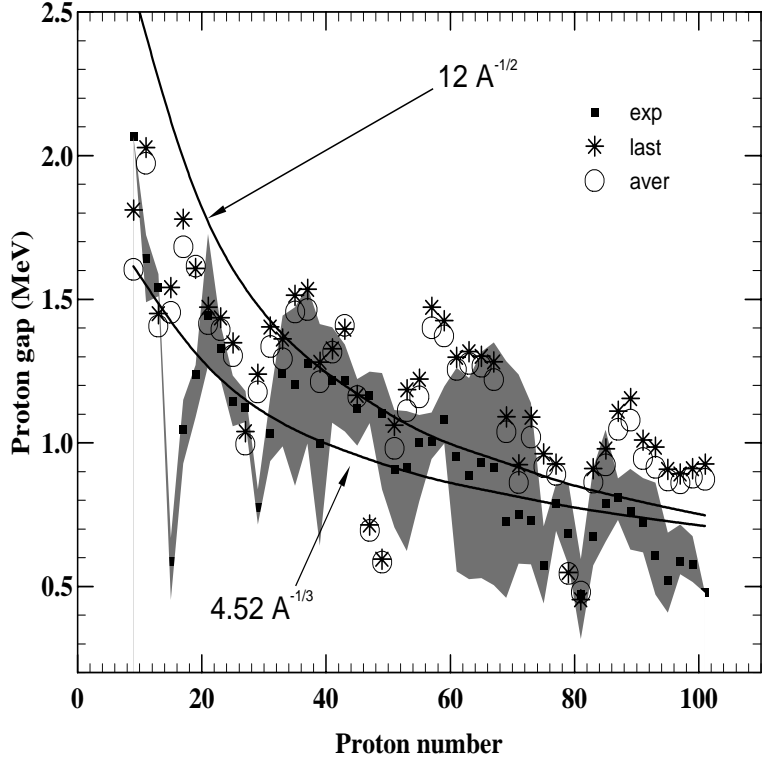}}}
\mbox{}
\caption{Comparison between experimental and theoretical proton pairing 
gaps plotted as functions of the proton number $Z$. Squares represent
experimental gaps extracted with the filter $\Delta^{(3)}$. Stars and circles
are the corresponding theoretical values $\overline{\Delta^{\pi}_{last}}$ and 
$\overline{\Delta^{\pi}_{aver}}$ defined in the text. The shaded area 
represents the gap limits between which the experimental data are found
before average. The lower curve corresponds to a least square fit on
experimental data imposing an $A^{-1/3}$ law.}
\label{fig1}
\end{figure}

\begin{figure}
\centerline{\hbox{\psfig{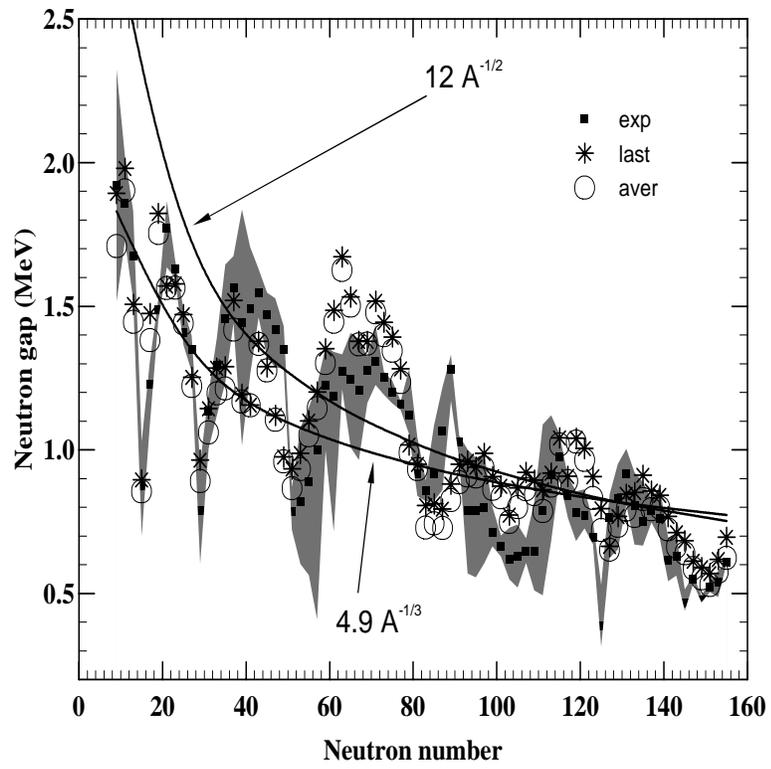}}}
\caption{Same as Fig.1 for neutrons.}
\label{fig2}
\end{figure}

 The theoretical gaps given by Eq.($\ref{avera}$) are displayed in
 Figs.~\ref{fig1} and ~\ref{fig2} together with the corresponding experimental
 data. It is important to mention here that even if theoretical calculations
 have been performed for even-even nuclei, we have deliberately plotted the
 gaps as function of the odd-$Z$ (resp. odd-$N$) values (the same as those
 used in ~\cite{Sat98} to extract experimental data) from which the
 neighboring even-even nuclei studied theoretically have been selected.

\begin{figure}
\centerline{\hbox{\psfig{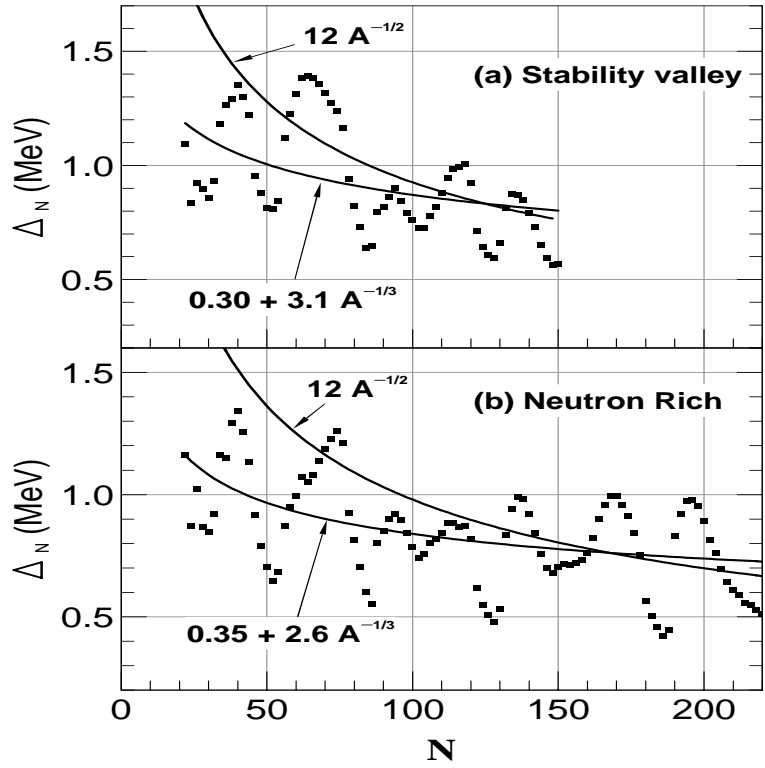}}}
\caption{Theoretical gaps for neutrons in the SV and NR regions. Note that
 zero gap values at magic numbers are included but lifted to finite values
 after averaging.}
\label{fig3}
\end{figure}

 In view of the great sensitivity of the gap with respect to all input
 parameters (force, effective mass, etc ...) there is an excellent overall
 agreement of the theoretical quantities with experiment. The $A^{-1/3}$-law 
 in the fits of Figs.~\ref{fig1} and ~\ref{fig2} has no particular deep 
 theoretical fundation (see, however the remarks made in connection with 
 Fig.~\ref{fig3}) and other A-dependences can represent the average trend as
 well. This average behavior makes it however clear that the A-dependence of
 the gaps is much weaker than the $\Delta = 12 \, A^{-1/2}\,$ MeV law
 previously assumed. This finding is very satisfying as these 
 theoretical results give further credit to the analysis of experimental 
 data in \cite{Sat98} also concluding that the A dependence of the gap is
 weaker than the $\Delta = 12 \, A^{-1/2}\,$ MeV law.

 For magic numbers, theoretical gaps go to zero, since there is no pairing. In
 this case, the $\Delta^{(3)}$ value deduced from experiment does not really
 describe a pairing effect, but rather an average of single-particle gaps
 around shell closures. Theoretical $\Delta$'s agree particularly well with
 experimental ones in mid-shell nuclei where experimental $\Delta^{(3)}$'s
 represent a genuine pairing effect. For these reasons, we did not include
 in the theoretical average, the nuclei having a magic number of protons
 or neutrons.

 In Fig.~\ref{fig1}, one notices that theoretical $\Delta_{aver}$ overestimate
 experimental data. One reason is the absence, in our calculations, of the
 Coulomb Interaction in the pairing field since it would require too much
 computing time. However, we have checked for a couple of nuclei that including
 it reduces the gap-values by 100 to 200 keV, depending on the nucleus proton
 number, thus improving the agreement with the experimental data.

 Other sources of uncertainty may partly be accounted for through the
 effective force. This is likely to be the case for the recently debated
 influence of surface vibrations on nuclear pairing \cite{prlpair,jetp99}
 which is claimed to give a sizeable contribution to nuclear superfluidity.
 Since, however, the gap values calculated from the D1S Gogny force are
 quite realistic (see Figs.~\ref{fig1} and ~\ref{fig2}), it is justified to
 assume that the Gogny force accounts for such effects at least on the average.


 In view of the good agreement of experiment and theory found in
 Figs.~\ref{fig1} and ~\ref{fig2}, we further investigate the average
 trend of the theoretical Gogny-HFB gap values versus A. For this purpose,
 we define $\Delta_{N}$ for a given $N$ as an arithmetic average over
 several $Z$-values -- taken in an interval so that the nucleus $(Z,N)$
 belongs either to what we call the stability valley (SV) or the neutron
 rich region (NR) -- of the theoretical $\Delta_{aver}$. Noticing that the
 relation \cite{book} $Z_{s}=A/(1.98+0.0155 A^{2/3})$ almost perfectly
 defines the most stable nuclei, the SV and NR are defined as the region
 $0.94 Z_{s} \leq Z \leq 1.05 Z_{s}$ and $Z \leq 0.94 Z_{s}$, 
 respectively. In order to further smoothen the curves, we also average
 the mean $\Delta_{N}$'s together with the $\Delta_{N \pm 4}$'s and
 $\Delta_{N \pm 2}$'s. The width $\Delta N=8$ of this last average should
 be small enough not to affect significantly the mean trends. Finally, this
 procedure gives us the full black squares in Fig.~\ref{fig3}. It is important
 to mention that for nuclei close to drip lines, the method used to solve the
 HFB equations does not allow us to include continuum effects. In order to
 test the validity of our results for such nuclei, we have checked that their
 pairing properties are stable with respect to a large increase
 of the harmonic oscillator basis. This test consists in introducing quite a
 different representation of unbound orbitals and therefore the observed 
 stability indicates that our theoretical $\Delta$'s are not significantly
 sensitive to continuum effects.

 From the obtained curve in Fig.~\ref{fig3}(a), one again clearly sees that the
 old $\Delta = 12 \, A^{-1/2}\,$ MeV law strongly overestimates the
 average trend, at least for small $A$. We also inserted our least square
 fit assuming a $\Delta_{N}=\alpha + \beta A^{-1/3}$ law. Justification for
 this choice stems from the weak coupling approximation for the gap, i.e.
 $\Delta \propto \exp(-1/G.\rho)$, where $G \propto 1/A$ is the usual
 constant pairing matrix element and $\rho \propto A(1+cA^{-1/3})$ the
 level density at the Fermi energy \cite{bohr}. Indeed, performing a
 Taylor-expansion in powers of the small parameter $c$ yields the above
 mentioned law for $\Delta$. It is also worth mentionning that such a mass 
 dependence of the pairing gap has also been obtained in ref.~\cite{vogel}
 (see also ref~\cite{moller}). The best fit values for Fig.~\ref{fig3}(a) are
 found to be $\alpha=0.3$ and $\beta=3.1$. Calculation indicate slightly
 different trends for $\Delta_{N}$ in SV and NR regions. In particular for
 Fig.~\ref{fig3}(b) we find $\alpha=0.35$ and $\beta=2.6$. The asymptotic
 value is rather close to the nuclear matter value $\Delta_{nm}=0.4$ MeV
 obtained with the Gogny Force \cite{pinf}.


 The fact that in Fig.~\ref{fig3}(b)
 rather large $\Delta_{N}$ values for large $N$ are found is likely an 
 indication of the increasing role of the neutron skin. Similar
 tendencies are obtained for the proton gaps (not shown). The different 
 average trends seen in Figs.~\ref{fig2} and ~\ref{fig3} (in particular for
 low $N$ values) are due to the fact that in Fig.~\ref{fig2} all experimentally
 available data are taken into account irrelevant whether they correspond to
 stable or exotic nuclei whereas in Fig.~\ref{fig3} two regions have been
 sorted out. Our choice $\Delta_{N}=\alpha+\beta A^{-1/3}$ is certainly not
 unique but the $A^{-1/3}$ dependence, besides having some theoretical
 justification as explained above, yields overall the best results among the 
 various choices we tried. For example, an improved fit with different
 parametrisation can be obtained in Fig.~\ref{fig3}(a), but there is no point
 in making a separate fit for each figure.

 The increase of the gap with decreasing size of the
 nucleus may eventually be a rather generic feature in meso- and nano-scopic
 systems. Indeed, also in small superconducting metallic grains and in thin
 superconducting films there seems to be a tendency for increasing
 gap-values as the size of the system is reduced \cite{last}. Whether the
 physical origin of the effect is the same in all cases remains to be seen.

 In summary we investigated the mass dependence of the average gap values for
 neutrons and protons in large scale HFB calculations with the Gogny D1S
 effective interaction. Very good agreement with the experimental filter
 $\Delta^{(3)}(N=2n+1)$ is found. This indicator was advocated previously
 \cite{Sat98} for its capability to eliminate spurious mean field components
 from the gap values in an optimal way. The present theoretical study
 therefore supports the much weaker dependence of the gap, advanced in
 \cite{Sat98}, than the so far accepted $\Delta = 12 \, A^{-1/2}\,$ MeV law.
 The agreement between experimental and theoretical size dependence of $\Delta$
 is a non trivial fact and this study may open similar investigations in other
 finite superfluid or superconducting systems.

 \mbox{}

 This work was supported by the Swedish Institute (SI) and the Polish
 Committee for Scientific Research (KBN) under contract No. 5 P03B01421
 and performed in the frame of the
 "Groupement de Recherche - Structure des Noyaux Exotiques". One
 of us (P.S.) acknowledges useful discussions with F. Hekking.

\newpage

\mbox{}

\newpage

\mbox{}

\newpage

\begin{thebibliography}{00}

\bibitem{braun} F. Braun et al., Phys. Rev. Lett. 79 (1997) 921.

\bibitem{marco} B. DeMarco and D. S. Jin, Science 285 (1999) 1703.

\bibitem{bohr} A. Bohr and B. R. Mottelson, Nuclear Structure, Vol I, Benjamin, New York (1969).

\bibitem{Dec80} J. Decharge and D. Gogny, Phys. Rev. C 21 (1980) 1568.

\bibitem{cpc91} J.F. Berger et al., Comp. Phys. Comm. 63 (1991) 365.

\bibitem{Sat98} W. Satu{\l}a et al., Phys. Rev. Lett. 81 (1998) 3599.

\bibitem{Jah37} H.A. Jahn and E. Teller, Proc. Roy. Soc. A161 (1937) 220.

\bibitem{Zeldes} N. Zeldes et al., Mat. Fys. Skr. Dan. Vid. Selsk. 3 (1967) No. 5.

\bibitem{Jen84} A.S. Jensen et al., Nucl. Phys. A 431 (1984) 393.

\bibitem{Man94} M. Manninen et al., Z. Phys. D 31 (1994) 259.

\bibitem{Dob} J. Dobaczewski et al., Phys. Rev. C 63 (2001) 024308.

\bibitem{gatl}  W. Satu{\l}a, AIP Conference Proceedings 481, ed. C. Baktash,
                p. 141; nucl-th/0003019.

\bibitem{Bender} M. Bender et al. Eur. Phys. J. A8 (2000) 59.

\bibitem{Dug1} T. Duguet et al. nucl-th/0105049 v2

\bibitem{Dug2} T. Duguet et al. nucl-th/0105050 v1

\bibitem{bogo} N.N.Bogoliubov, Sov. Phys. JETP 7 (1958) 41.

\bibitem{BloMes} C.Bloch and A.Messiah, Nucl. Phys. 39 (1962) 95.

\bibitem{girod} M. Girod and B. Grammaticos, Phys. Rev. C 27 (1983) 2317.

\bibitem{prlpair} F. Barranco et al., Phys. Rev. Lett. 83 (1999) 2147.

\bibitem{jetp99} A. V. Avdeenkov and S. P. Kamerdzhiev, JETP Lett. 69 (1999) 715.

\bibitem{book}  P. Marnier and E. Sheldon, Physics of Nuclei and Particles, Part I, Academic Press Inc, London (1969), p.36.

\bibitem{vogel} P. Vogel et al., Phys. Lett. B 139 (1984) 227.

\bibitem{moller} P. M\"{o}ller and J. R. Nix, Nucl. Phys. A 520 (1990) 369c.

\bibitem{pinf} H. Kucharek et al., Phys. Lett. B 216 (1989) 249.

\bibitem{last} C. T. Black et al., Phys. Rev. Lett. 76 (1996) 688, and references therein.

\end{thebibliography}
\end{document}